# Imaging the breakdown and restoration of topological protection in magnetic topological insulator MnBi$_2$Te$_4$

Qile Li[1,2*], Iolanda Di Bernardo[1,2], Johnathon Maniatis[1], Daniel McEwen[1,2], Liam Watson[1,2], Benjamin Lowe[1,2], Thi-Hai-Yen Vu[1], Chi Xuan Trang[1,2], Jinwoong Hwang[3,4], Sung-Kwan Mo[3], Michael S. Fuhrer[1,2], Mark T. Edmonds[1,2,5*]

[1]School of Physics and Astronomy, Monash University, Clayton, VIC, Australia

[2]ARC Centre for Future Low Energy Electronics Technologies, Monash University, Clayton, VIC, Australia

[3]Advanced Light Source, Lawrence Berkeley National Laboratory, Berkeley, CA, 94720 USA

[4]Department of Physics and Institute of Quantum Convergence Technology, Kangwon National University, Chuncheon, 24341, Republic of Korea

[5]ANFF-VIC Technology Fellow, Melbourne Centre for Nanofabrication, Victorian Node of the Australian National Fabrication Facility, Clayton, VIC 3168, Australia

*Corresponding Author mark.edmonds@monash.edu and qile.li@monash.edu

**Abstract –**

**Quantum anomalous Hall (QAH) insulators transport charge without resistance along topologically protected chiral one-dimensional edge states. Yet, in magnetic topological insulators (MTI) to date, topological protection is far from robust, with the zero-magnetic field QAH effect only realised at temperatures an order of magnitude below the Néel temperature $T_N$, though small magnetic fields can stabilize QAH effect. Understanding why topological protection breaks down is therefore essential to realising QAH effect at higher temperatures. Here we use a scanning tunnelling microscope to directly map the size of the exchange gap ($E_{g,ex}$) and its spatial fluctuation in the QAH insulator 5-layer MnBi$_2$Te$_4$. We observe long-range fluctuations of $E_{g,ex}$ with values ranging between 0 (gapless) and 70 meV, uncorrelated to individual point defects. We directly image the breakdown of topological protection, showing**

**that the chiral edge state, the hallmark signature of a QAH insulator, hybridizes with extended gapless metallic regions in the bulk. Finally, we unambiguously demonstrate that the gapless regions originate in magnetic disorder, by demonstrating that a small magnetic field restores $E_{g,ex}$ in these regions, explaining the recovery of topological protection in magnetic fields. Our results indicate that overcoming magnetic disorder is key to exploiting the unique properties of QAH insulators.**

**Main Text-**

Topological protection has become a crucial concept in the recent development of condensed matter physics[1-4]. In the quantized versions of the Hall effect (QHE), spin Hall effect (QSHE) and anomalous Hall effect (QAHE), topological protection manifests as one-dimensional electronic edge states where scattering due to local perturbations is prohibited[5]. This opens the way towards high-temperature lossless electronic transport applications[6,7] as well as new approaches to topologically-protected fault-tolerant quantum computing[8,9]. These technologies require robust topologically protected edge channels, but in electronic devices this protection is often observed to break down. Breakdown of the QHE due to disorder, temperature, and current has been understood within scaling theory[10], and was a fundamental development in the understanding of continuous quantum phase transitions[11]. The microscopic origins of disorder-induced QHE breakdown are still a vibrant area of investigation, with new developments in graphene showing unique aspects of backscattering in the presence of both electron- and hole-like edge channels[12,13] The QSHE is robust to non-magnetic disorder at zero temperature[14], but magnetic disorder can cause scattering at finite temperatures[15], with a quantum phase transition from helical liquid to insulator under strong interactions[14]. In contrast, the chiral quantum anomalous Hall (QAH) edge channel supports unidirectional flow of electrical current that should be robust to any potential perturbations smaller than the exchange energy gap $E_{g,ex}$ which is opened in the surface-state of a thin film 3D topological insulator (TI) via long-range magnetic order.[16,17] Despite this, breakdown of topological protection is ubiquitously observed at far lower temperatures than $E_{g,ex}/k_B$ or $T_N$, where $k_B$ is the Boltzmann constant. In dilute magnetic doped topological insulators, this is thought to be due to the magnetic diorder, leading to non-uniform magnetization and a fragile QAHE that is only observable at extremely low temperatures (<1 K).[18,19]

Intrinsic stoichiometric magnetic topological insulators (MTIs) which possess both non-trivial topology and intrinsic magnetism, for example $MnBi_2Te_4$[20,21] should in principle circumvent the issues associated with dilute magnetic doping. Promisingly, odd-layers of $MnBi_2Te_4$ in the 2D limit host quantum anomalous Hall states [22], leading to observation of the QAHE at 1.4 K, with the temperature increasing to 6.5 K under application of an external magnetic field[23]. Yet, this value is still substantially lower than $T_N = 25$ K and the activation energy extracted from transport measurements of $\Delta E = 0.64$ meV is two orders of magnitude smaller than the predicted $E_{g,ex} = 70$ meV (800K). Furthermore, QAHE is not routinely observed in ultra-thin odd-layer $MnBi_2Te_4$ samples or quantization is only observable in a large perpendicular magnetic field[24]. These results hint at the presence of various types of surface disorder that act to suppress the $E_{g,ex}$[25] and destroy QAHE.

To understand the mechanism of topological breakdown requires direct measurement of the interplay between surface disorder, local fluctuations in $E_{g,ex}$, and the chiral edge state with atomic-scale precision using low-temperature scanning tunneling microscopy and spectroscopy (STM/STS). A technique previously used to probe band gap fluctuations and edge states in other 2D materials[2,18,28-31]. To date, most STM/STS measurements on $MnBi_2Te_4$ have been performed on bulk crystals and have focused on point defects[26-28]. Little attention has been paid to ultra-thin films of $MnBi_2Te_4$ and the mechanisms of topological protection breakdown and suppression of QAHE. A recent report suggests connection between local magnetic $Mn_{Bi}$, $Bi_{Mn}$ anti-site defects (notation $X_Y$ means a X ion replaces a Y ion in the lattice) and collapse of the Dirac mass gap in high defect regions, but did not measure bandgap fluctuations over large areas to understand possible short-range behavior from magnetic disorder, how disorder interacts with the chiral edge state [29] or how the disorder effects respond to a magnetic field. Thus, the mechanism by which topological protection is destroyed, as well as how it recovers in $B$ field, are still not understood.

In this work, we utilize magnetic field-dependent STM/STS to study the origin of QAHE suppression in 5 septuple layer (SL) MnBi$_2$Te$_4$. We directly measure spatial fluctuation of $E_{g,ex}$ and importantly observe the electronic overlap of the chiral edge state with gapless metallic bulk states, providing the route to breakdown of the QAHE. Finally, we demonstrate that by applying a magnetic field well below the spin-flop transition, we are able to restore the magnetic gap in the gapless regions, explaining the recovery of QAHE in small magnetic fields.

**Fig. 1(a)-(b)** presents the crystal structure of one septuple layer (SL) MnBi$_2$Te$_4$. Lattice constants, magnetic moments and possible lattice defects are labelled. Within each SL, intra-layer Mn$^{2+}$ ions are coupled through ferromagnetic interaction. Between two adjacent SLs, two Mn$^{2+}$ atomic layers are coupled through anti-ferromagnetic (AFM) interaction, resulting in thickness-dependent magnetic properties. We grow high-quality epitaxial ultra-thin MnBi$_2$Te$_4$ using molecular beam epitaxy (MBE) on $p$-type Si(111)–7×7 to achieve MnBi$_2$Te$_4$ films where the Fermi level resides in the surface gap (see Methods for growth details, and structural characterization in Supplementary **S1**). Supplementary **Fig. S1(b)** shows a typical large-area STM topography scan with coexisting regions of 4 and 5 SL MnBi$_2$Te$_4$ islands that are atomically flat, along with small pinholes of bare substrate. **Fig. 1(c)** shows a 20×20 nm atomic resolution STM image revealing the expected 1×1 atomic surface structure with lattice constant 4.3 Å. Several different defects are present; bright spots on the surface correspond to negatively charged Bi$_{Te}$ point defects whilst the dark triangles are Mn$_{Bi}$ defects, similar defects have been observed in Cr, Mn-doped 3D TIs[19,30,31]. The third defect type – $Bi_{Mn}$ (located in the middle of each SL) is not directly visible in (c) but presented in **Supplementary Fig. S2**.

**Fig. 1(d)** shows an ARPES spectrum of 5 SL $MnBi_2Te_4$ thin film grown on *n*-type Si(111) taken at 8 K along ΓM direction where a Dirac cone is clearly visible near the Fermi level. The strong spectral weight near Γ in the Dirac point region could be due to Te-orbital-related matrix elements[32,33] or be the result of bandgap fluctuations as the spectral signal is averaged over the beam spot size (100×100 μm). To demonstrate this possibility we fit the ARPES spectrum in **Fig. 1(d)** with three possible scenarios (red curve: full band gap $E_{g,ex}$ = 70 meV extracted from energy distribution curve analysis[32], green curve: reduced $E_{g,ex}$ = 45 meV, blue curve: gapless $E_{g,ex}$ = 0 meV limit).

To properly understand the $E_{g,ex}$ distribution, we use STS to measure the d$I$/d$V$ spectrum (the differential conductance d$I$/d$V$ as a function of sample bias $V$) which is proportional to the local density of states (LDOS) at energy $E_F + eV$, where $e$ is the elementary charge. **Fig. 1(e)** shows three typical STS from the same SL terrace. The size of $E_{g,ex}$ changes drastically with location: the red STS curve corresponds to $E_{g,ex}$ = 70 meV, the green curve shows a reduced band gap $E_{g,ex}$ = 45 meV, and the blue curve is consistent with a gapless ($E_{g,ex}$ = 0) spectrum (see **Fig. S3** and **Fig. S4** for details on extracting bandgap and the minimal tip-induced band bending). All spectra in **Fig. 1(e)** were taken more than 5 nm away from step edges in order to exclude effects from edge states.

Before turning to the origin of the band gap variations, we confirm that our 5 SL $MnBi_2Te_4$ is indeed a QAHI by probing the step edge between 4 and 5 SL $MnBi_2Te_4$ (shown in **Fig. 2(a)**), to verify the presence of the conductive edge state, a consequence of topological protection and signature of a QAHI. Since the chiral edge state (CES) exists within the Dirac band gap, the sample bias was tuned into the Dirac band gap to image the edge state. **Fig. 2(b)** shows a d$I$/d$V$ map taken at +25 meV: a pronounced increase in d$I$/d$V$ signal is observed that is localized at the step edge, indicating a conductive edge state. For a perfect QAHI there should be a well-defined suppression of the bulk LDOS within the bandgap, but we observe bulk regions well away from the edge that also show strong LDOS at the same energy, indicating the presence of disordered bulk metallic regions directly connected to the edge state. To confirm that some disordered bulk regions are indeed metallic, in **Fig. 2(c)** we measure d$I$/d$V$ spectra corresponding to the edge state (red curve), disordered bulk states (cyan curve) and normal bulk states (purple curve). Spectra positions are marked in **Fig. 2(a)** and **(b)**. The normal bulk region away from the edge shows the expected insulating behavior with $E_{g,ex}$ = 60 meV, but the d$I$/d$V$ spectra at both the edge and within these disordered bulk state regions are quite different, with states filling the entire bulk gap, indicating a continuous metallic percolative path for electron transport from the edges through the bulk. **Figure 2(d), (e)** shows spatial d$I$/d$V$ profiles as a function of distance away from the edge $x$ measured along the two red dashed lines in **(a)-(b)** (labelled as Cut 1 and Cut 2). In **Fig. 2(d)**, the edge state is localized on the edge around $x$ = 4 nm and appears as a sharp streak, as marked by the red arrow. The other streak around $x$ = 6 nm is clearly separated from the edge state and we assign it to disordered bulk states, marked by a white arrow. In **Fig. 2(e)**, the edge state is less prominent and several streaks with periodic spacing (white arrows) are prominent corresponding to disordered bulk states. This suggests the edge state hybridizes with the metallic regions formed by these disordered bulk states, and these metallic regions represent continuous conductive pathways that guide the CES into the bulk, leading to the conductive breakdown of QAHE through dissipative bulk conduction and resulting in non-perfect quantization of Hall conductance and non-zero longitudinal resistance.

To investigate the origin of these metallic regions formed by disordered bulk states, we perform atomic resolution topography and STS maps around the magnetic defects Mn$_{Bi}$ and Bi$_{Mn}$, which allow us to extract maps of the spatial variation of $E_{g,ex}$, and the gap center energy $E_c$ to determine the influence each defect has on the electronic structure. $E_c$ is equivalent to extrapolating the massive Dirac bands linearly into the gap to obtain the Dirac point in the gapless limit. At locations where the Dirac bands are gapped, $E_c$ is a good measure of local doping shifts associated to the magnetic order. $Bi_{Te}$ defects, are non-magnetic, thus, are unlikely to result in fluctuations in the exchange energy gap $E_{g,ex}$. **Figure 3(a)** illustrates the position of a $Mn_{Bi}$ defect in the crystal lattice, and **Fig. 3(b)** shows an atomic resolution image of 10×5 nm area (-500 mV, 3 nA) with $Mn_{Bi}$ defects marked in black triangles. The substitution of $Bi^{3+}$ by $Mn^{2+}$ causes contraction on the three neighboring surface Te atoms[27]. **Fig. 3(c)-(d)** are maps of $E_{g,ex}$ and $E_c$ extracted from the d$I$/d$V$ spectra on the same area in **(b)** (details found in **Supplementary Fig. S3**). Representative d$I$/d$V$ curves from different locations, marked with $E_{g,ex}$ and $E_c$ values, are plotted in **Supplementary Fig. S5**. The $Mn_{Bi}$ defects (green triangles) have little contribution to the fluctuations in $E_{g,ex}$ and $E_c$ in **(c)** and **(d)** respectively, with only small decrease in $E_{g,ex}$ and slight increase in *n*-type doping (due to the negative charge of the $Mn_{Bi}$ defects) observed. Instead, the fluctuation appears to be dominated by some underlying structure on the scale of a few nanometers, causing $E_{g,ex}$ and $E_c$ to be correlated. Gap size histograms of $Mn_{Bi}$ regions and $Mn_{Bi}$ excluded regions are shown in **Fig. 3(e)**. Overall, the patterns of gapped and gapless regions formed over several nanometers are very different from the band gap fluctuations in dilute magnetic doped TIs[19, 31], suggesting the presence of magnetic disorder. In **Fig. 3(f)**, a $Bi_{Mn}$ defect in the middle atomic layer is depicted, where the substitution results in a non-magnetic defect. Such defects manifest as large bright triangles and are only visible at positive bias as marked by purple triangles in **(g)**. The three bright dots in each triangle are due to Te p-orbitals on the surface responding to $Bi_{Mn}$ defects[27]. d$I$/d$V$ mapping was performed in the area marked by the yellow box in **(g),** with the band gap and gap center maps shown in **Fig. 3(h)-(i)**. The band gap map in **(h)** and the histogram in **(j)** show that $Bi_{Mn}$ defects are not magnetic, resulting in gapless regions. However, regions well away from the $Bi_{Mn}$ defect still display band gap fluctuation with significant weight of gapless states, suggesting that $Bi_{Mn}$ defects alone do not result in extended regions of suppressed Dirac band gap and metallicity in the bulk. The results in **Fig. 3** therefore demonstrate the band gap and gap center fluctuations cannot be explained entirely by local suppression by any of the three types of isolated point defects discussed above, and imply the possibility that longer-ranged collective behavior of disorder is responsible for the extended Dirac band gap suppression on the surface of MnBi$_2$Te$_4$.

To understand the origin of the observed large-scale band gap fluctuations, we measure magnetic field-dependent STS to examine how the extended gap suppression structures respond to $B_\perp$ field. **Figure 4(a)** shows STM topography of the atomically flat scan region at $B_\perp$=0 T**.** In **Fig. 4(b)** STS curves taken at different locations show three types of behavior: gapless regions (blue), gapped regions with fluctuating Dirac band gap (green and red) and regions where the Dirac electron band is suppressed and manifests as an anomalously large bulk gap (black). Such conduction band (CB) suppression has been previously observed in bulk MnBi$_2$Te$_4$[27]. The diminished CB intensity prevents us from extracting accurate values of $E_{g,ex}$, thus, the suppressed regions are masked in black in the following gap maps and excluded in subsequent analysis. **Figure 4(d)** plots STS curves taken at the same location (green circle in **(a)**) that is initially gapless at $B_\perp$=0 T (blue curve) and $B_\perp$=1 T (red curve). It is immediately clear that a 1 T field is sufficient to restore $E_{g,ex}$ to 40 meV with enhanced exchange coupling. Having observed magnetic field induced band gap modulation, we now perform d$I$/d$V$ mapping (-150 mV, 0.4 nA) on an 80×80 point-mesh on the same 30× 30 nm area in **(a)** at $B_\perp$=0 T (**Fig. 4(c)**) and $B_\perp$=1 T (**Fig. 4(e)**). Histograms of $E_{g,ex}$ with and without $B$ field are shown in **Fig. 4(f)**. These maps reflect the spatial fluctuation of $E_{g,ex}$ over larger scale and will be used to investigate its origin beyond point defects. The histogram in the upper panel of **Fig. 4(f)**, shows prominent weighting for $E_{g,ex}$ < 10 meV, corresponding to a skewed distribution (skewness 0.91) with mean of 26.3 meV and standard deviation of 25.8 meV. Upon applying $B_\perp$=1 T, the histogram in the lower panel of **Fig. 4(f)** shows a significant reduction in $E_{g,ex}$ < 10 meV regions, and a gap opening and renormalization that results in a near-normal distribution (skewness 0.06) with an increased mean of 44.3 meV and smaller standard deviation of 20.2 meV.

We now consider possible origins of the extended suppressed gap structures. As recently observed in magnetic force microscopy measurements[34], whilst the bulk of MnBi$_2$Te$_4$ thin film remains AFM coupled, the surface exhibits magnetic spin flops which could be enhanced by Bi$_{Mn}$ defects or changes to the size of the inter-layer van der Waals gap near the surface[35]. In our thin film MnBi$_2$Te$_4$ samples, we are able to align the magnetic moments at $B_\perp$=1 T, much lower than required for inducing surface (2-3.5 T) and bulk spin flop (7.7 T) in previous work[34,36,37], which indicates that surface spin flop has little contribution to the exchange gap fluctuation observed. This suggests that there is significant magnetic disorder in the first SL that causes band gap fluctuation on the nanometer scale. Such magnetic disorder occurs in the Mn$^{2+}$ layer located at the center of the top few SLs, which is very hard to probe using STM. However, the exchange interaction of the Dirac states with Mn$^{2+}$ ions enables indirect mapping of such magnetic disorder based on the suppression of $E_{g,ex}$ as seen in **Fig. 4(c)** and **(e)**.

**Fig. 4(g)** illustrates the situation schematically. Magnetic disorder causes local suppression of the exchange gap (blue gapless Dirac spectrum) in extended regions due to long-range exchange interactions among local moments, while some regions retain partial (green band structure) or full (red band structure) gaps. Application of $B_\perp$=1T (bottom) aligns the moments of magnetically disordered regions, increasing the gap (red band structure). Interestingly, the $B$ field also decreases the area of suppressed CB regions, suggesting CB suppression is also related to disordered magnetic moments beyond the influence of deficient Bi orbitals due to Mn$_{Bi}$ defects[27]. Finally, we propose an explanation to the origin of the magnetic disorder. The prevalent band gap fluctuation observed implies weakened inter-layer and intra-layer exchange interaction in 5 SL MnBi$_2$Te$_4$. Its low magnetic anisotropy energy makes MnBi$_2$Te$_4$ similar to a 2D Heisenberg magnet that does not sustain long-range ferromagnetic order[37,38]. Such weakened magnetic anisotropy makes the magnetic ordering more vulnerable to magnetic defects, especially $Bi_{Mn}$. With $Mn^{2+}$ ions replaced by non-magnetic $Bi^{3+}$ ions, exchange coupling between intra-layer $Mn^{2+}$ ions is weakened, depending on the concentration of such defects. Therefore, the ferromagnetic configuration of $Mn^{2+}$ ions could be canted and disordered in the presence of a large amount of $Bi_{Mn}$ defects resulting in reduced magnetization and gapless spectra over extended areas, as predicted recently[29]. This is indeed shown in **Supplementary Fig. S6** on an area that contains a larger number of $Bi_{Mn}$ defects, and is indeed mostly gapless. Upon applying the 1T field, a significant reduction of gapless regions occurs (further details in **Supplementary Fig. S6**).

Using magnetic field STM/STS measurements, we have demonstrated that the chiral edge state in QAHI 5 SL $MnBi_2Te_4$ is directly coupled to extended percolating bulk metallic regions arising from band gap fluctuation caused by magnetic surface disorder. By applying a magnetic field, the band gap fluctuations can be greatly reduced, and the average exchange gap increased to 44 meV, close to predicted values for 5 SL $MnBi_2Te_4$[21,22,32]. These results provide insight on the mechanism of topological breakdown and how it can be restored in a magnetic field[23,24]. Minimizing magnetic disorder will be the key to realizing QAHE in not only odd-layer $MnBi_2Te_4$ ultra-thin films but also other MTIs at elevated temperature in the future. The weak magnetic anisotropy in $MnBi_2Te_4$ makes it difficult to sustain long-range magnetic order, and improved $MnBi_2Te_4$ crystal or film growth alone may not be sufficient to fully mitigate magnetic disorder. Therefore, other strategies such as heterostructure engineering $MnBi_2Te_4$ with other robust, highly anisotropic 2D ferromagnets [39] or ferromagnetic/topological insulators sandwich heterostructures [40-42] may be required to achieve the robust topological protection required for next-generation lossless electronics and topological quantum computing[6-9].

## Methods

**Growth of Ultra-thin MnBi$_2$Te$_4$ on Si(111)**

Ultra-thin MnBi$_2$Te$_4$ thin films were grown in a Scienta Omicron Lab 10 molecular beam epitaxy (MBE) growth chamber. The Si(111) substrate was flash-annealed at 1180°C with direct current heating to achieve an atomically flat (7×7) surface reconstruction. Effusion cells were used to evaporate elemental Mn (99.9%), Bi (99.999%) and Te (99.95%). A quartz crystal microbalance was used to calibrate rates before growth and reflection high-energy electron diffraction (RHEED) was used to monitor the crystal growth in-situ. Each SL of MnBi$_2$Te$_4$ was grown by first growing 1 quintuple-layer Bi$_2$Te$_3$ followed by growing a bilayer MnTe in overflux of Te at 230 °C. 1 SL MnBi$_2$Te$_4$ forms spontaneously by re-arranging MnTe layer into the middle of 1QL Bi$_2$Te$_3$ similar to MnBi$_2$Se$_4$[43]. The growth time for each 1QL Bi$_2$Te$_3$ and MnTe was calibrated from the oscillation of the RHEED pattern. Then the process was repeated five times to reach the desired thickness and finished with a post-annealing process in Te flux for 10 min to improve crystallinity. The films were subsequently capped with 10nm amorphous Te, to allow transfer in air to the STM chamber.

**Scanning Tunneling Microscopy/Spectroscopy (STM/STS) Measurements**

The capped films were transferred in air to a Createc LT-STM chamber and were annealed in UHV at 290°C for 2.5 hours to remove the Te capping before performing STM measurements at 4.3 K. A PtIr tip was prepared and calibrated using an Au (111) single crystal, confirming the presence of the Shockley surface state at -0.5 V and flat LDOS near the Fermi level before all measurements. The STM differential conduction measurements (d$I$/d$V$) were performed using standard lock-in method with 8 mV AC excitation voltage at 797Hz . Differential conductance measurements were made under open feedback conditions with the tip in a fixed position above the surface. For the magnetic field dependent STM/STS measurements, a 1T magnetic field was applied perpendicular to the sample.

**Angle-resolved Photoemission Spectroscopy (ARPES) Measurements**

ARPES measurements were performed at Beamline 10.0.1 at Advanced Light Source (ALS) in Lawrence Berkeley National Laboratory, USA. A 5 SL MnBi$_2$Te$_4$ sample was grown following the same growth procedure in an MBE system integrated with the beamline endstation and transferred into the measurement chamber after growth. Data was taken using a Scienta R4000 analyser at 8K and photon energy of 50 eV was selected to optimize the signal. The combined energy resolution is 15-20 meV and the angular resolution is 0.2°, or equivalent to $0.01 Å^{-1}$ momentum resolution for the photon energy used.

**Acknowledgements**

M. T. E., Q. L., M. S. F., I. D. B. acknowledge funding support from ARC Centre for Future Low Energy Electronics Technologies (FLEET) CE170100039. Q. Li acknowledges funding support from the AINSE postgraduate award. This research used resources of the Advanced Light Source, which is a DOE Office and Science User Facility under contract no. DE-AC02-05CH11231. M.T.E. and Q.L. acknowledge travel funding provided by the International Synchrotron Access Program (ISAP) managed by the Australian Synchrotron, part of ANSTO, and funded by the Australian Government. This work was performed in part at the Melbourne Centre for Nanofabrication (MCN), the Victorian Node of the Australian National Fabrication Facility (ANFF).

**Author contributions**

M. T. E and Q. L. devised the STM experiments. Q. L. performed the MBE growth and STM/STS measurements at Monash University. I. D. B., B. L., L. W. and T.H.Y.V assisted the scanning probe measurements. J. H., S.-K. M. and C. X. T assisted the ARPES measurements. Q. L. performed data analysis with assistance from M. T. E, M. S. F, J. M and D. M. Q. Li, and M. T. E. composed the manuscript. All authors read and contributed feedback to the manuscript.

# FIGURES

FIGURE 1

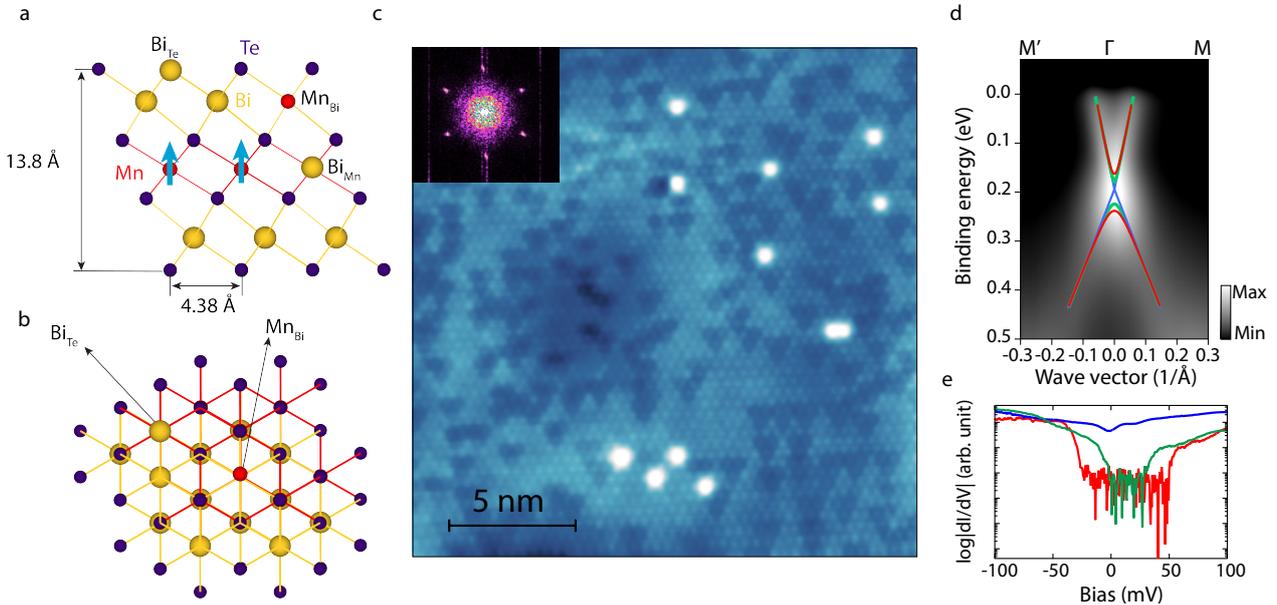

**Figure 1 | Characterization of epitaxial ultra-thin MnBi$_2$Te$_4$ and overall electronic structure from ARPES. a, b** Crystal structures of a septuple layer of MnBi$_2$Te$_4$. **(a)** side view of the lattice with lattice constants, atom species, and defects labelled. The magnetic moments on Mn$^{2+}$ ions are marked with blue arrows. **(b)** Top view of the lattice. **(c)** Atomic resolution image (-2 V, 180 pA) of a flat 20×20 nm area where Mn$_{Bi}$ (dark triangles) and Bi$_{Te}$ (bright dots) defects are clearly visible. The insert shows the fast Fourier transformed image of the same area. (Note the spots corresponding to 1×1 surface atomic structure). **(d)** Angle-resolved photoemission spectrum of five-layer MnBi$_2$Te$_4$ along Γ-M where the fully gapped ($E_{g,ex}$ = 70 meV) band dispersion is marked by red curve. Green and blue are illustrations of possible reduced gap and gapless dispersions ($E_{g,ex}$ = 35 meV and 0 meV respectively). **(e)** d$I$/d$V$ spectra taken at different locations on the same terrace of 5 SL MnBi$_2$Te$_4$ (0.2 V, 400 pA) showing gapless (blue curve), reduced gap (green curve) and fully gapped d$I$/d$V$ curves from different regions on the same terrace.

FIGURE 2

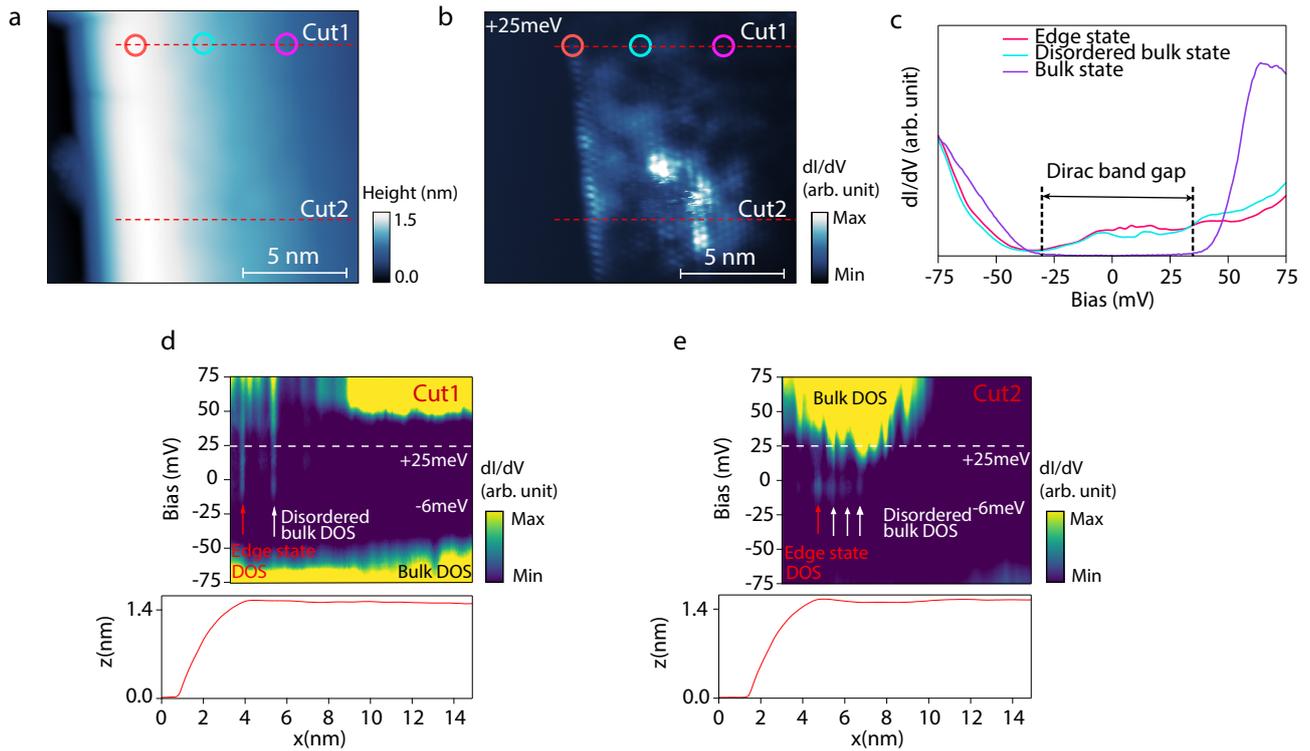

**Figure 2 | Visualizing the QAHE edge state and its coupling to bulk metallic states. (a)** Topography image (-1 V, 50 pA) taken across a step edge from 4 SL to 5 SL MnBi$_2$Te$_4$. **(b)** d$I$/d$V$ map at +25 mV bias in the same region as (a) to show the spatial distribution of the edge state. **(c)** d$I$/d$V$ spectra taken from edge state region (red circle), disordered bulk region (cyan circle) and normal bulk region (purple circle) as marked in **(a)** and **(b)**. **(d),(e)** d$I$/d$V$ spectra and height profile taken across the edge from top of the area and bottom of the area respectively marked by red dashed lines. The horizontal axis of the spectra is aligned with the height profile. The edge state is marked by red arrow and other peaks are attributed to disordered bulk states which is marked by white arrows. The white horizontal dashed line shows the bias at which **(b)** is taken.

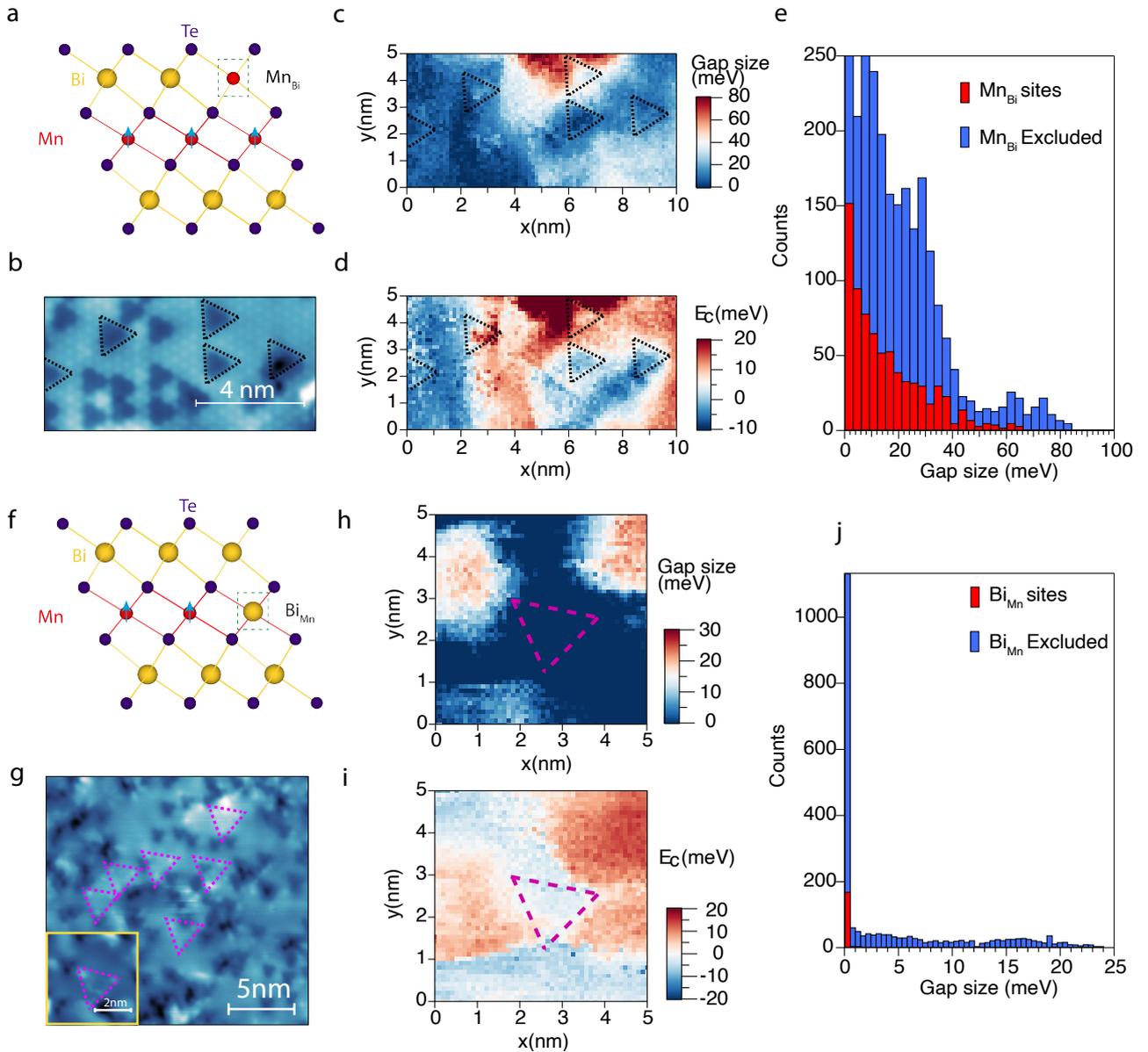

**Figure 3 | Local response of the exchange gap and doping to point defects. (a)** Illustration of a $Mn_{Bi}$ defect in the lattice. **(b)** Topography of a 10×5 nm area (-150 mV, 3 nA) with $Mn_{Bi}$ defects manifesting as dark triangles (marked in black triangles). **(c)** An exchange gap, $E_{g,ex}$, map extracted from d$I$/d$V$ spectra (-100 mV, 0.8 nA) on a 40×80 mesh for visualizing band gap fluctuation and **(d)** gap center from the same region as **(c)**. **(e)** Histograms of the $E_{g,ex}$ extracted from regions with and without $Mn_{Bi}$ defects respectively. **(f)** Illustration of a $Bi_{Mn}$ defect in the lattice. **(g)** Topography of a 40×40 nm area (+1.7 eV, 80 pA) with $Bi_{Mn}$ defects which manifest as bigger bright triangles (marked as purple triangles). Insert: a 5×5 nm region where d$I$/d$V$ spectra (-100 mV, 0.91 nA) on a 50×50 mesh was taken to show its effect on $E_{g,ex}$ **(h)** and doping level which is reflected on gap center **(i)**. **(j)** Histograms of $E_{g,ex}$ extracted from the defect region and region excluding the defect.

FIGURE 4

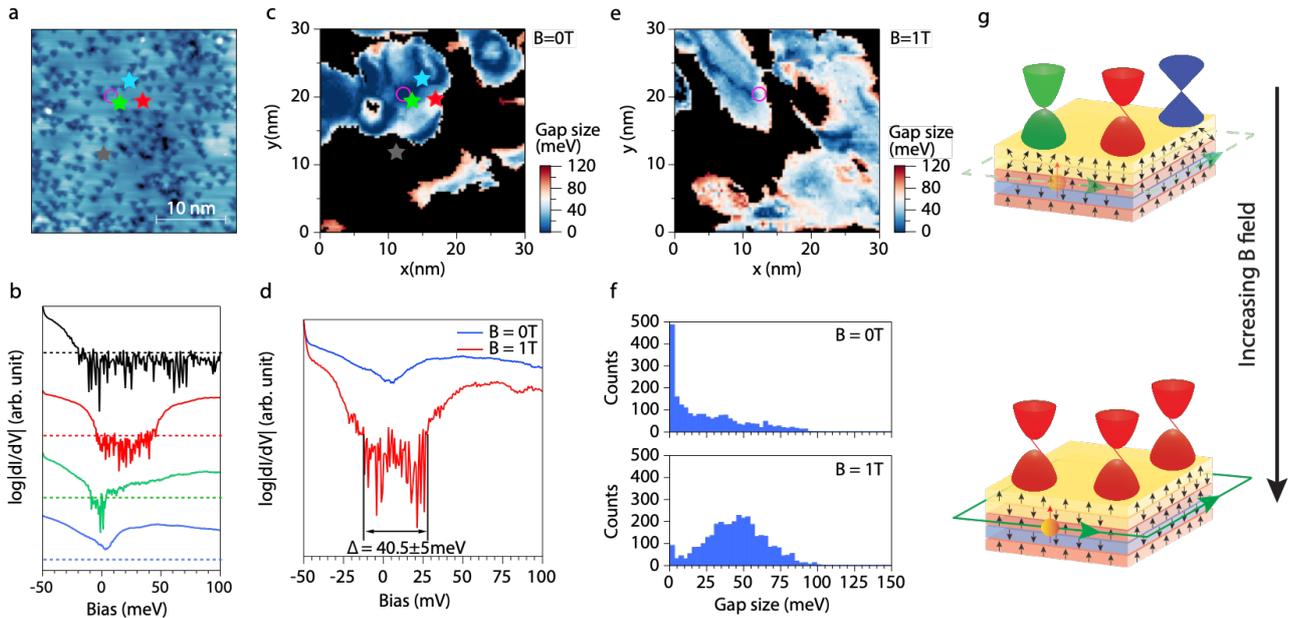

**Figure 4 | Magnetic field-induced modulation of the exchange gap. (a)** Topography scan (-0.5 V, 100 pA) of a 30×30 nm area where magnetic field dependent STS measurements were conducted. **(b)** Representative d$I$/d$V$ spectra taken at different locations, blue: gapless regions, green: reduced-gap regions, red: large-gap regions and black: regions where Dirac electron band is suppressed which prevents us from extracting the band gap. **(c)** Band gap map (80×80 points, -150 mV, 0.4 nA) of the region in **(a)** at magnetic field $B$ = 0 T. **(d)** d$I$/d$V$ spectrum taken at the position marked by purple circle in **(a)** at $B$ = 0 T (blue) and $B$ = 1 T (red). A magnetic Dirac gap of 40.5 meV is opened with 1 T field in a gapless region at 0 T. **(e)** Band gap map (80×80 points) of the same region in (a) at $B$ = 1 T. Black regions in (c) and (e) correspond to the suppressed Dirac electron band regions which prevent accurate determination of the exchange gap and are excluded from the maps and subsequent histograms in (f). **(f)** Histograms showing gap size at $B$ = 0 T (upper panel) and $B$ = 1 T (lower panel). A clear renormalization of bandgap is observed with magnetic field. **(g)** Illustration of the band gap spatial fluctuation caused by surface magnetic disorder which can be reduced significantly by applying a perpendicular magnetic field. The blue, green and red Dirac cones represent gapless, partially gapped and fully gapped regions. Their representative d$I$/d$V$ curves can be found in **Figure 1(f)**. Upon applying a perpendicular magnetic field, the exchange gap in the Dirac cones increases until it reaches saturation.

Supporting Information

# Imaging the breakdown and restoration of topological protection in magnetic topological insulator MnBi$_2$Te$_4$

Qile Li[1,2], Iolanda Di Bernardo[1,2], Johnathon Maniatis[1], Daniel McEwen[1,2], Liam Watson[1,2], Benjamin Lowe[1,2], Thi-Hai-Yen Vu[1], Chi Xuan Trang[1,2], Jinwoong Hwang[3,4], Sung-Kwan Mo[3], Michael S. Fuhrer[1,2], Mark T. Edmonds[1,2,5]*

[1]School of Physics and Astronomy, Monash University, Clayton, VIC, Australia

[2]ARC Centre for Future Low Energy Electronics Technologies, Monash University, Clayton, VIC, Australia

[3]Advanced Light Source, Lawrence Berkeley National Laboratory, Berkeley, CA, 94720 USA

[4]Department of Physics and Institute of Quantum Convergence Technology, Kangwon National University, Chuncheon, 24341, Republic of Korea

[5]ANFF-VIC Technology Fellow, Melbourne Centre for Nanofabrication, Victorian Node of the Australian National Fabrication Facility, Clayton, VIC 3168, Australia

*Corresponding Author mark.edmonds@monash.edu and qile.li@monash.edu

**Table of Contents**

1. Structure and surface characterization of 5 SL MnBi$_2$Te$_4$ epitaxial film
2. Schematics and determination of Bi$_{Mn}$ defects
3. Determination of Dirac band gap and gap center from d$I$/d$V$ spectrum
4. Bias dependent and set point current dependent STS measurements
5. Representative d$I$/d$V$ spectra from STS map around Mn$_{Bi}$ defects
6. B-field dependent Dirac band gap maps showing band gap fluctuation from region with large number of Bi$_{Mn}$ defects

1. Structure and surface characterization of 5 SL MnBi$_2$Te$_4$ epitaxial film

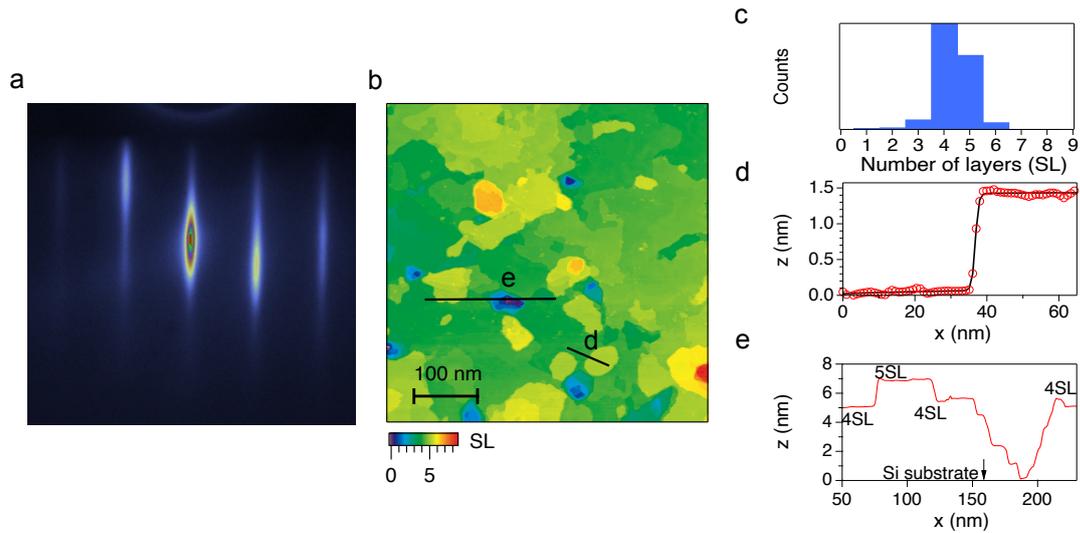

**Figure S1 | Structure and surface characterization of ultra-thin MnBi$_2$Te$_4$ epitaxial film. (a)** Reflection high energy electron diffraction of the thin film showing strong sharp streaks which indicates high crystallinity. **(b)** STM topography scan on a 500×500 nm area (-2 V, 20 pA). **(c)** Histogram of number of SLs from **(b)** which shows majority of 4 SL terrace (green) with regions of 5 SL terrace (yellow). The thickness is determined from the depth of pinholes (dark blue and purple regions) which represent the Si substrate. **(d)** A line profile extracted from one of the 4 SL-5 SL step edges as marked by the black line in **(b)** and fitting to an edge function, which yields a step edge of 1.39 nm. **(e)** A line profile extracted across a pin hole to show the thickness of terraces as marked by the black line in **(b)**.

2. Schematics and determination of Bi$_{Mn}$ defects

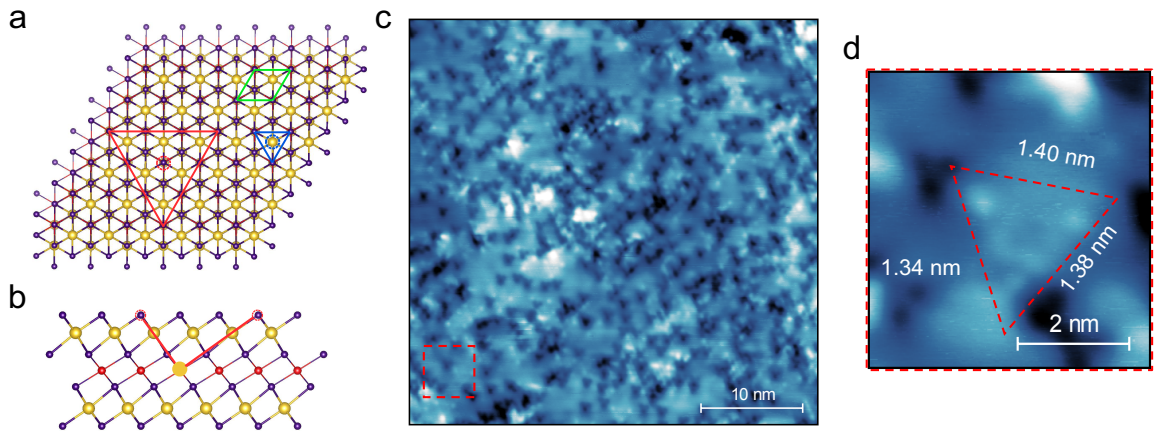

**Figure S2 | Schematics and determination of Bi$_{Mn}$ defects. (a)** Top view of the crystal structure with primitive cell (green), Mn$_{Bi}$ defect (blue) and Bi$_{Mn}$ defect (red) marked. **(b)** Side view of the Bi$_{Mn}$ defect in a SL. The Te atoms on the very top appear as bright triangles at positive bias because of DOS propagated along Te $p_z$ orbitals centered around Bi$_{Mn}$ defect. The three Te atoms appear to be more positively charged due to the extra charge from a Bi$^{3+}$ ion replacing Mn$^{2+}$ ion. **(c)** Topography scan on a 40×40 nm area (+1.7 V, 80 pA) showing Bi$_{Mn}$ defects on an atomically flat terrace. **(d)** A zoom-in image of the selected region marked by red box in **(c)** for extracting dimensions of defects. The length of the triangle's edge is around 1.35 nm which is close to the theoretical value of 1.31 nm. The same orientation of Mn$_{Bi}$ dark triangles and Bi$_{Mn}$ bright triangles also matches the schematics **(a)**.

3. Determination of Dirac band gap and gap center from d$I$/d$V$ spectrum

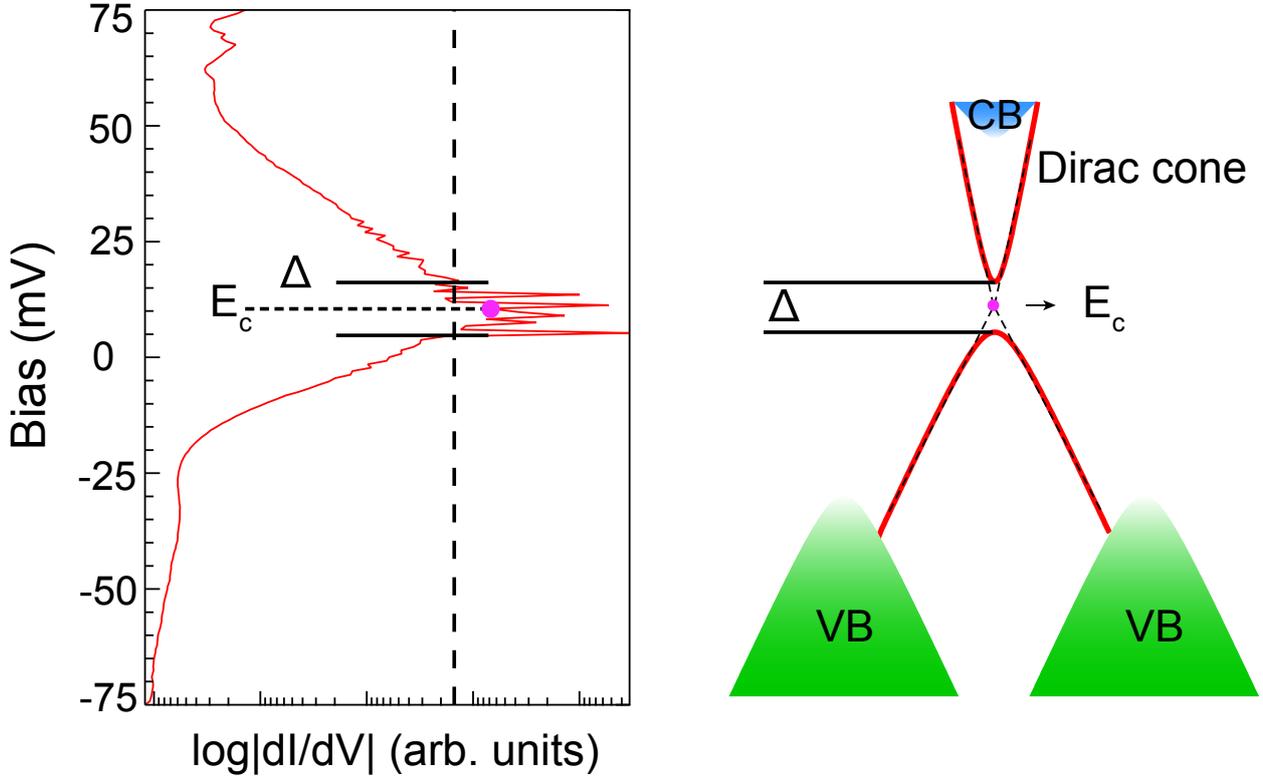

**Figure S3 | Determination of Dirac band gap and gap center from *dI/dV* spectrum. Left**, a d$I$/d$V$ spectrum taken on a 5 SL MnBi$_2$Te$_4$ terrace showing a gap in the Dirac states. **Right,** a schematic of the Dirac cone corresponding to the d$I$/d$V$ spectrum on the left. The Dirac cone in 5 SL MnBi$_2$Te$_4$ is lifted out of VB and resides between bulk conduction band (CB) and bulk valence band (VB). The Dirac gap is extracted based on the noise floor of the d$I$/d$V$ spectrum (black dashed line). The width of the region between valence and conduction band edges corresponds to the size of Dirac gap Δ as illustrated on the right figure.

To extract the Dirac gap from STS spectra, the edges of valence band and conduction band are determined from the spectrum as the onset of d$I$/d$V$ intensity above the noise floor (black dashed line in Figure S3 left). A logarithmic scale is chosen for d$I$/d$V$ intensity axis to better account for the sudden change of intensity near the band edge. The local doping level can be estimated from the center of Dirac gap which can be calculated using

$$E_c = \frac{\int f(\mathbf{r}, E) E dE}{\int f(\mathbf{r}, E) dE}$$

where $f(\mathbf{r}, E)=1$ if STS curve $g(\mathbf{r}, E)$ < noise floor and $f(\mathbf{r}, E)=0$ otherwise[19]. This is essentially averaging the position of center of gap and is equivalent to extrapolating the gapped Dirac bands as shown in Figure S3 right. Since the Dirac cone is now gapped by magnetic order and Dirac point no longer exists, the gap center (purple dot) is a good measure of the local doping shift by defects.

## 4. Bias dependent and set point current dependent STS measurements

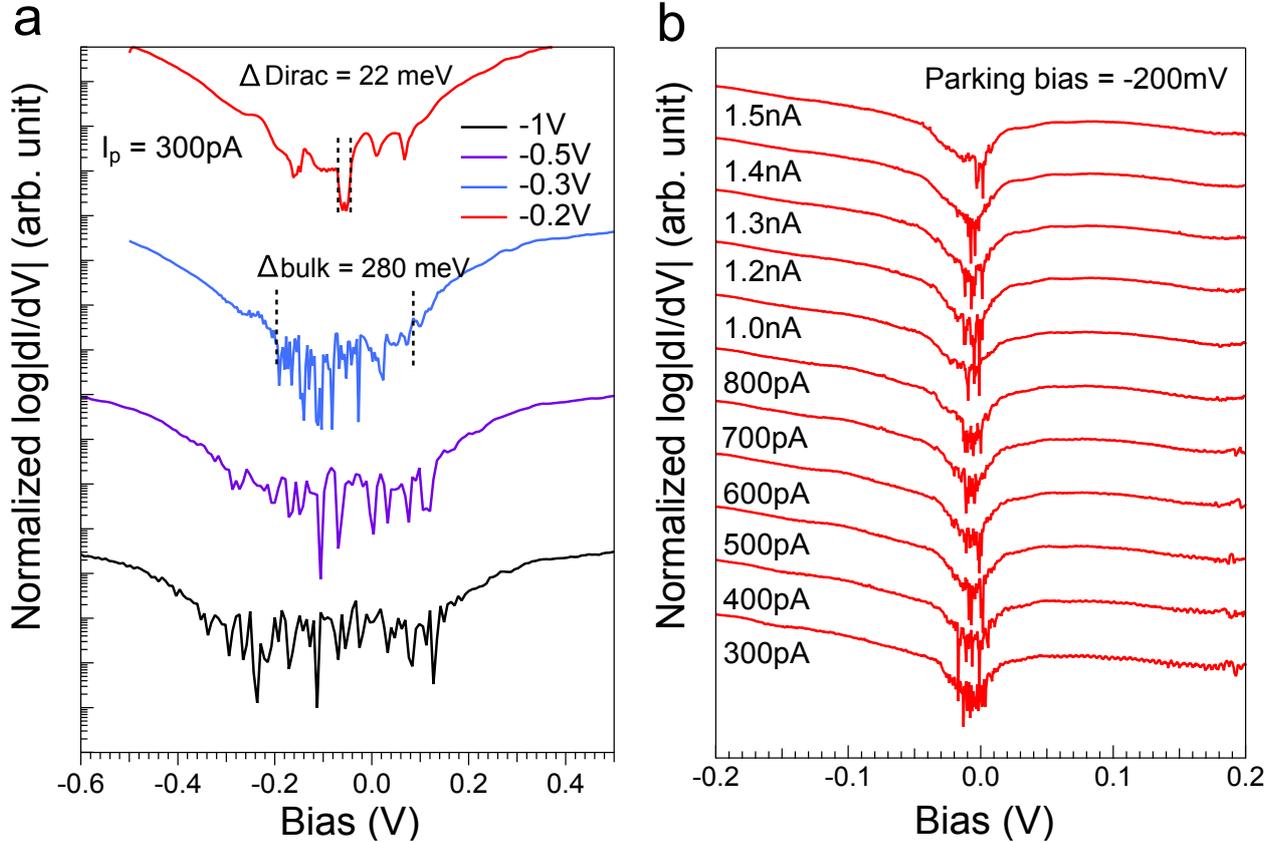

**Figure S4 | Bias dependent and set point current dependent STS measurements. (a)** STS spectra taken at different bias, the tip-sample tunneling junction decreases as the bias voltage decreases bias, resulting in a stronger signal which is necessary to observe the Dirac bands in dI/dV. **(b)** STS spectra taken at various set point current and fixed parking bias of -200 mV showing a Dirac gap. The band edge position shows minimal shift at different set point current and bias, indicating minimal tip induced band bending.

As shown in **Figure S4(a)**, the STS spectrum on the terrace only shows a large bulk gap at parking bias of -1 V. As the tip is parked closer to the surface by decreasing bias, more features in the spectrum, including surface states and Dirac band gap, can be resolved within the bulk gap. Therefore, we choose parking bias of typical value -0.2 V and 300 pA for our STS maps. In **Figure S4(b)**, STS spectra taken on the same location and fixed bias but with varying set point current are plotted and offset manually. From set point current of 300 pA to 1.5 nA, there is no significant shift of band edges or increase of band gap. In the case of tip induced band bending, increasing the set point current will increase the band bending and lead to increase of band gap. Apparently, the most noticeable change in the **Figure S4(b)** is reduced noise in the spectra while the overall shape and position of the band edge stay the unchanged. Therefore, we can rule out contribution from tip induced band bending to the band gap.

## 5. Representative d*I*/d*V* spectra from STS map around Mn$_{Bi}$ defects

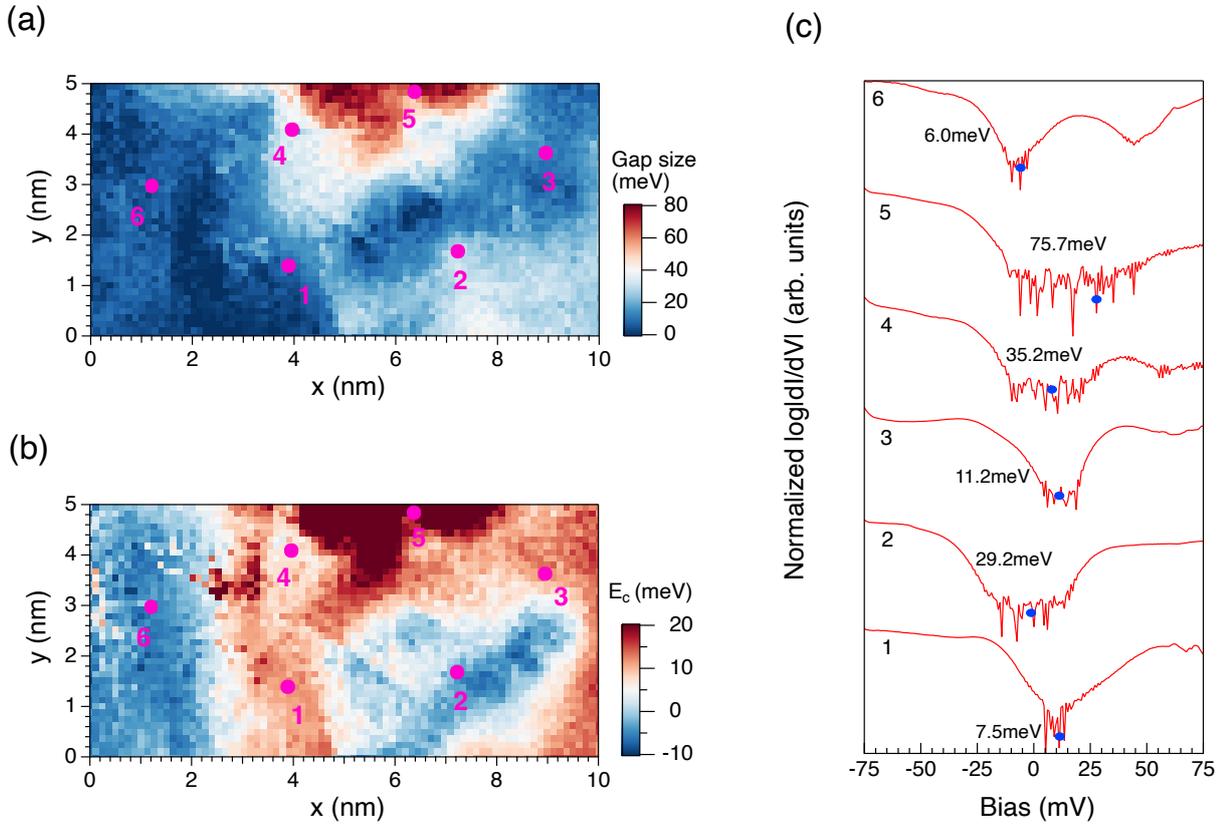

**Figure S5 | Representative d*I*/d*V* spectra from STS map around Mn$_{Bi}$ defects. (a)** Band gap map extracted from d*I*/d*V* spectra with some locations where STS shown marked in green. **(b)** Gap center map extracted from d*I*/d*V* spectra. **(c)** Stack plot of d*I*/d*V* spectra taken from locations in **(a)-(b)**, where band gap values are marked on each d*I*/d*V* spectrum, and gap center positions calculated using method discussed in **Figure S3** are marked by blue points.

## 6. B-field dependent Dirac band gap maps showing band gap fluctuation from region with large number of $Bi_{Mn}$ defects

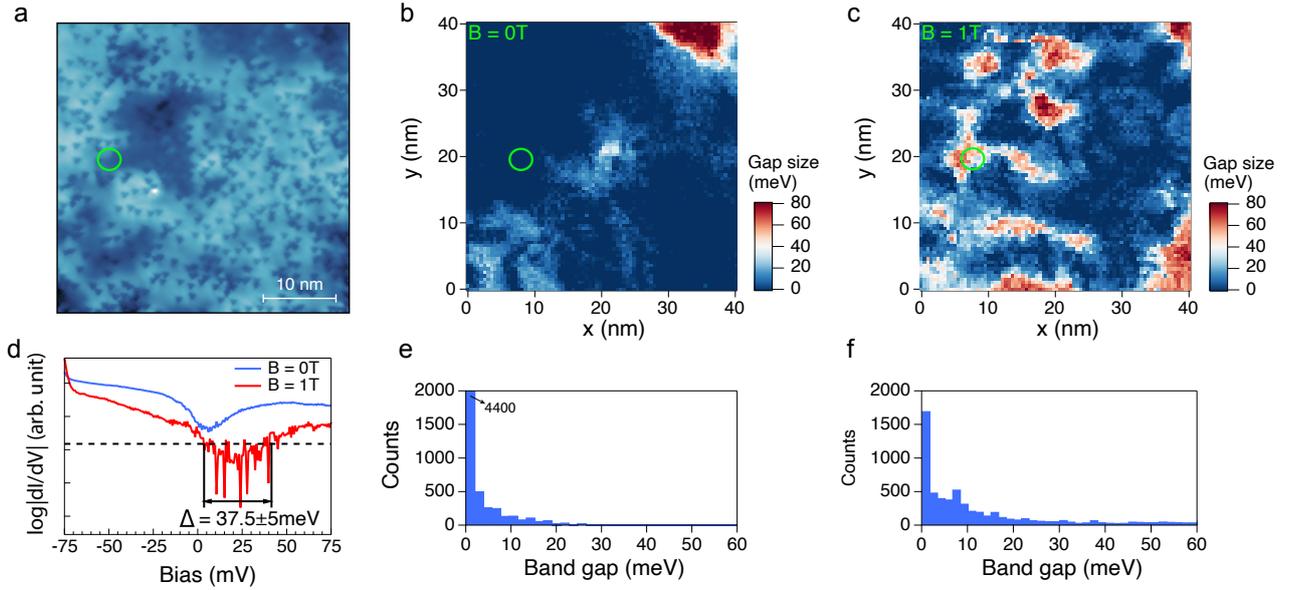

**Figure S6 | B-field dependent Dirac band gap maps showing band gap fluctuation from region with $Bi_{Mn}$ defects.** (a) Topography scan of the same area as in Figure SF2 (80 pA, -1 V). (b) d$I$/d$V$ spectra taken from the green circle in (a) at B=0 T (blue) and 1T (red). An exchange gap is opened in 1 T field where exchange coupling is enhanced by the external field. (c) Spatial dependent of Dirac band gap modulation at B=0 T. (d) Band gap histogram extracted from (c) which shows significant counts from gapless regions (dark red in c). (e) Dirac band gap map taken at 1 T field where gapped regions now form pattern (blue). (f) Histogram extracted from (e) which shows that Dirac band gaps are increased with a drastic decrease of gapless regions.

**Figure S6** shows magnetic field dependent Dirac band gap maps from a region with a large concentration of $Bi_{Mn}$ defects. Clearly, upon applying a perpendicular magnetic field of 1 T, the band gap map shows up more regions with small band gap around 10 meV. In some regions the band gap has increased from gapless to moderate value of 40 meV. As shown in (f), there is now visible counts from regions with gap larger than 30 meV. Statistical analysis on the maps in (b) and (c) shows an increase of average band gap value from 4.6 meV to 14.8 meV. The emerging counts from moderate band gap region results in an increase of standard deviation from 12.5 meV to 18.3 meV and similar to the results in **Figure 4**, the restoration of Dirac band gap is accompanied by decreasing skewness from 4.5 to 1.5. The results in **Figure S6** indicates that regions with large numbers of $Bi_{Mn}$ defects are typically more much more gapless and the exchange gap can be restored partially with a perpendicular magnetic field of 1 T. Because $Bi_{Mn}$ defects are non-magnetic, the pattern emerged in **Figure S6c** reflects the change of magnetic disorder in the center $Mn^{2+}$ layer. The amount of $Bi_{Mn}$ defects (see **Figure S2c**) could be responsible for the magnetic disorder.